# Stability and Migration of Small Copper Clusters in Amorphous Dielectrics

David M. Guzman, Nicolas Onofrio, and Alejandro Strachan[1]

*School of Materials Engineering and Birck Nanotechnology Center,*

*Purdue University, West Lafayette, Indiana, USA*

We use density functional theory (DFT) to study the thermodynamic stability and migration of copper ions and small clusters embedded in amorphous silicon dioxide. We perform the calculations over an ensemble of statistically independent structures to quantify the role of the intrinsic atomic-level variability in the amorphous matrix affect the properties. The predicted formation energy of a Cu ion in the silica matrix is 2.7±2.4 eV, significantly lower the value for crystalline $SiO_2$. Interestingly, we find that Cu clusters of any size are energetically favorable as compared to isolated ions; showing that the formation of metallic clusters does not require overcoming a nucleation barrier as is often assumed. We also find a broad distribution of activation energies for Cu migration, from 0.4 to 1.1 eV. This study provides insights into the stability of nanoscale metallic clusters in silica of interest in electrochemical metallization cell memories and optoelectronics.

---

[1] Corresponding author: strachan@purdue.edu

## I. INTRODUCTION

The dissolution and clustering of metallic ions in dielectric materials is important for a wide range of applications including microelectronics where metallic interconnects can dissolve into its surrounding dielectric [1-4], photonics where metallic nanoclusters are used to engineer optical properties [1, 5], and electrochemical metallization (ECM) cells for resistive switching random access memory (RRAM). RRAM devices operate by switching between low and high resistance states and ECM cells are promising candidates for next-generation non-volatile memory due to the potential of high scalability, fast switching speeds and ultra low power operation [6-10]. ECM cells consist of an electro active electrode (typically Cu or Ag) and an electro-inactive electrode separated by a dielectric like $SiO_2$ [8, 11] or solid electrolyte such as GeS [12] or GeSe [13]. The resistive switching mechanism is based on the dissolution of the active electrode into the dielectric under the application of a voltage of the appropriate polarity, the migration of the metallic ions through the dielectric toward the inactive electrode, the formation of small clusters and, eventually, the formation of a conductive metallic bridge that short-circuits the electrodes [14]. Thus, the low resistance state is achieved by the electrochemical formation of a metallic filament (a process called SET), while the electrochemical dissolution of the filament results in a high resistance state (RESET).

In this paper we study the energetics of dissolution and clustering of Cu ions in amorphous $SiO_2$ as well as their mobility using a combination of molecular dynamics with reactive interatomic potentials and density functional theory. The $Cu/SiO_2$ system is of significant interest for ECM cell applications [7, 15-17] due to desirable properties such as bipolar switching [11] and high compatibility with back-end-of-line processing in CMOS integrated circuits[18]. Both experimental studies [19-21] and atomistic simulations [14] have shown the formation of small, stable clusters to be the rate-limiting step in the switching on nanoscale ECM cells. Reactive MD simulations showed that the SET process involves the dissolution of nanoscale filaments and the formation of new clusters that are stabilized by reduction. Recent *in situ* electron microscopy studies revealed the electrochemical-driven dynamics of small metallic clusters and the results used to explain switching behavior [22]. However, the energetics and kinetics that govern this process remain unknown; for example, does the formation of Cu clusters involve nucleation and growth as postulated in Ref.[20] and prior work? If so, what is the critical nucleus size and surface energy? Our *ab initio* results show that the formation of clusters is always energetically favorable and does not require overcoming a nucleation barrier if the process involves long timescales that allow the silica atoms to relax around the Cu cluster. In addition we find relatively low activation barriers for Cu diffusion in amorphous silica (in the range 0.4 to 1.1 eV); this finding supports cluster formation as the rate-limiting step.

The rest of the paper is organized as follows. Section II describes the MD and DFT simulations in detail; in Section III we discuss the formation and clusterization energy of Cu ions and clusters in amorphous silica. Section IV presents activation energies for diffusion of copper in amorphous silicon dioxide. Conclusions are provided in Section V.

## II. SIMULATION DETAILS

We report on the thermodynamic stability of non-interacting Cu ions and small clusters in amorphous $SiO_2$ based on first-principles density functional theory calculations. In order to generate the initial structures for the DFT calculations we annealed a set of independent liquid structures using MD with the ReaxFF reactive force field [23]. This procedure was shown to generate accurate amorphous structures of silica [24, 25] and silicon nitride [26]. The use of an ensemble of structures is critical to capture the intrinsic atomic-level variability in the amorphous matrix. We also studied the diffusion of Cu ions in amorphous silica using transition state theory based on the nudged elastic band method [27, 28]. The following sub-sections provide details of the simulations.

### A. Amorphous Structures Models

The procedure to generate structure starts with a 3 x 3 x 2 orthorhombic supercell of α-cristobalite $SiO_2$ from which the two central unit cells are removed to accommodate the copper atoms or clusters. The cells contain 64 $SiO_2$ formula units (192 atoms) and the number of Cu atoms was varied between 1 and 8. The atomic interactions are described with the ReaxFF force field [23] described in Ref. [14] and all MD simulations are carried out using LAMMPS [29].

The initial structures are heated to 4000 K (above the melting temperature) and thermalized for 0.5 ns under isothermal, isobaric conditions (NPT ensemble) using MD with a time step of 0.5 fs. In the liquid state the simulation cells are deformed to a cubic shape maintaining the cell volume constant. The liquid samples are then equilibrated for 1.5 ns under isothermal, isochoric conditions (NVT) and 50 statistically independent melt samples are extracted from these simulations. Each of these structures is subsequently annealed to 300K in 4.0 ns using a stepwise cooling scheme, resulting in well-relaxed, statistically independent, structures. These 50 relaxed MD geometries are subsequently used as initial structures for the DFT calculations, as discussed in the following subsection. Figure **1** shows representative structures amorphous $SiO_2$ structures with Cu clusters generated with the procedure described above.

For the generation of copper clusters embedded into the amorphous $SiO_2$ matrix we use fictitious bonds to keep the Cu atoms together as a cluster at high temperatures while the silica matrix is melted around it. In contrast, for the amorphous silica with non-interacting Cu ions no bond constraints are imposed on the Cu atoms and are allowed move freely within the $SiO_2$ matrix. We studied planar, linear (Figure 1 top two rows of snapshots) and equiaxed clusters (Figure 1 bottom panels). Prior DFT calculations in vacuum [30, 31] have shown planar structures to be energetically favorable up to a size of 6 atoms and we aim to quantify if this hold for Cu clusters dissolved in amorphous $SiO_2$. In addition to planar and equiaxed configurations, single-atom chains are attractive for ECM cell memories as the ultimately scaled filament. Similar to the copper clusters, the Cu chains are generated by imposing fictitious bonds between the atoms and constraining the bond angle to force the copper atoms to remain in a linear chain configuration during the melting process. For the last steps of equilibration at 300K all fictitious bonds and constrains are removed.

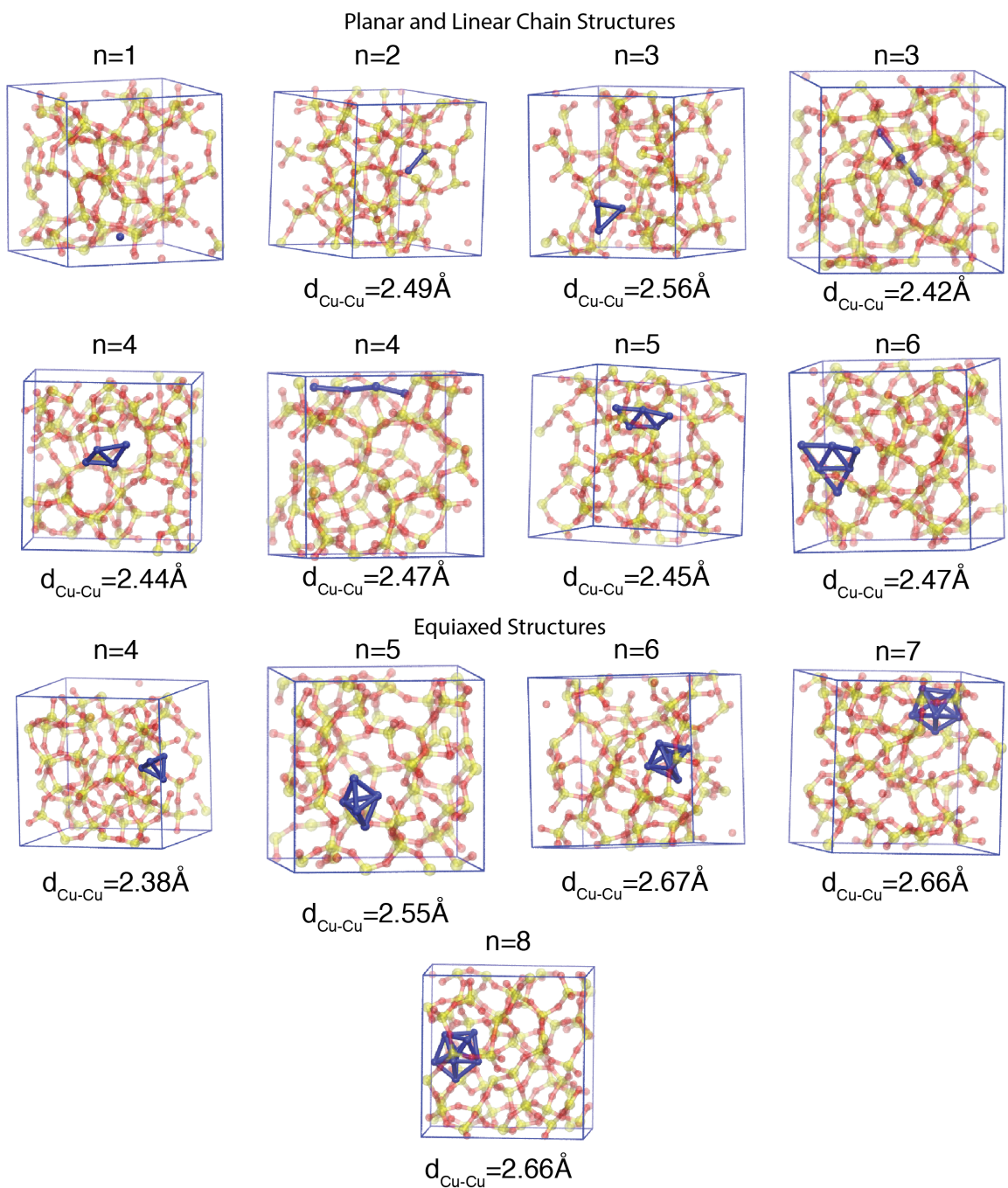

**Figure 1** Snapshots of relaxed Cu/SiO2 structures with clusters for 1 to 8 atoms. The structures of three- and four-atom linear chains and planar structures are also shown. These structures correspond to the final DFT relaxations after MD annealing using ReaxFF. $d_{Cu-Cu}$ is the average Cu-Cu bond.

## B. Density Functional Theory Calculations

The geometry optimization and electronic structure calculations were carried out using DFT as implemented in the Vienna *ab-initio* simulation package (VASP) [32, 33]. Projector-augmented-wave (PAW) potentials [34] were used to account for the electron-ion interactions, and the electron exchange-correlation potential was calculated using the generalized gradient approximation (GGA) within the Perdew-Burke-Ernzerhof (PBE) [35] scheme. The kinetic energy cutoff for all calculations was set to $500 eV$ and due to the relatively large simulation size only the gamma-point was used for the k-space sampling.

The structural relaxation of all structures was performed using a conjugate gradient (CG) algorithm with a force tolerance of $0.01 eV/Å$ and the electronic relaxation was carried out with a tolerance of $1\times10^{-5} eV$.

## C. Nudged Elastic Band Characterization of Migration

We used the nudged elastic band (NEB) method as implemented in VASP [28] to determine minimum barrier energy diffusion paths between known initial and final geometries representing metastable configurations of the Cu/SiO$_2$ system. The NEB calculations start from a set of geometries interpolating [36] between initial and final structures; then, the ionic positions of the different geometries are iteratively optimized using only the ionic-force components perpendicular to the hypertangent. The energy along the path is later determined by spline interpolation based on the total energy of the individual geometries and the tangential projection of the 3N force components on each geometry.

The computation of energy barriers via the NEB method in amorphous systems is not straightforward since the method requires a priori knowledge of initial and final configurations of the transition state chain. Due to the random nature of amorphous structures there is no direct way to determine metastable configurations of the Cu/SiO$_2$ system. In order to identify meaningful configurations to be used in the NEB calculations, we performed MD simulations at 800K of a copper ion diffusing in amorphous SiO$_2$ using ReaxFF and identified subsequent metastable configurations between Cu jumps. These structures were used as initial and final states for the NEB calculations.

## III. FORMATION AND CLUSTERIZATION ENERGIES

We define the formation energy of *n* Cu atoms dissolved in amorphous SiO$_2$ using fcc Cu and amorphous silica as reference states using:

$$E_{form} = E(Cu_n / SiO_2) - \left[ \langle E(SiO_2) \rangle + nE_{fcc}(Cu) \right],$$

where $E(Cu_n / SiO_2)$ is the total energy of the simulation with *n* Cu atoms dissolved in the silica matrix, $\langle E(SiO_2) \rangle$ is the average energy of amorphous SiO$_2$ samples with the same number of formula units as the system of interest, and $E_{fcc}(Cu)$ is the energy per atom of a fcc Cu unit cell.

Blue histograms in Figure 2 show the distribution of formation energies of equiaxed copper clusters with size 1 to 8 atoms. Interestingly, the distribution of formation energies is quite broad with standard deviations between 2.4 and 3.8 eV, see Table 1. This is not surprising, as defect formation energies in amorphous silica [25] show similar fluctuations and so does the cohesive energy of amorphous silica samples, shown for reference as red histograms in Figure 2 (with a mean value shifted to zero).

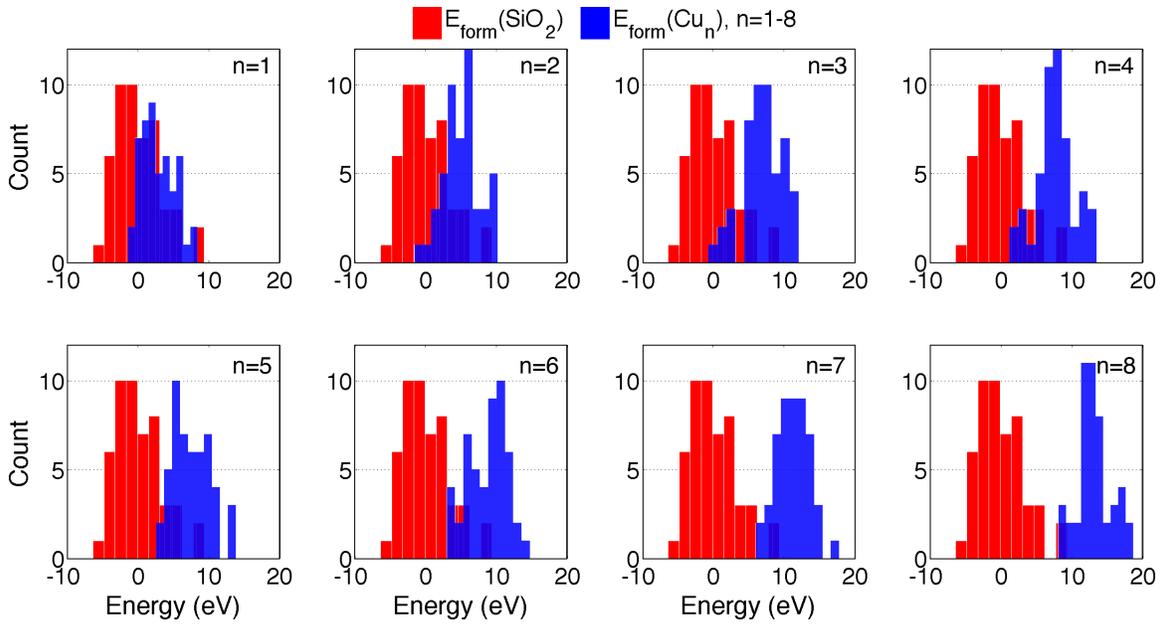

**Figure 2** Formation energy distributions of equiaxed Cu clusters 1-8 atoms in size. The red distributions correspond to the total energy distribution of a-SiO$_2$, which is the same for all cases. The blue distribution represents the energy of forming a Cu cluster of n atoms by implanting Cu atoms in a-SiO$_2$ from its most stable phase (FCC).

**Table 1** Average and standard deviation values of the formation energy and Bader electrons of copper ions and clusters embedded in amorphous silica.

| | Isolated Ions | | Clusters | |
|---|---|---|---|---|
| # Atoms | $E_{form}$ (eV) | Charge (e) | $E_{form}$ (eV) | Charge (e) |
| 1 | 2.7±2.4 | 0.59±0.17 | | |
| 2 | 5.8±2.4 | 0.55±0.18 | 5.1±2.5 | 0.42±0.19 |
| 3 | 7.8±2.1 | 0.50±0.12 | 6.9±2.8 | 0.37±0.11 |
| 4 | 11.6±3.8 | 0.49±0.09 | 7.6±2.7 | 0.31±0.07 |
| 5 | 14.7±2.5 | 0.48±0.10 | 8.6±2.8 | 0.26±0.09 |
| 6 | 17.4±3.0 | 0.48±0.06 | 10.7±2.7 | 0.25±0.07 |
| 7 | 19.1±3.1 | 0.47±0.05 | 11.1±2.4 | 0.22±0.06 |
| 8 | 21.7±2.9 | 0.46±0.08 | 13.0±2.4 | 0.21±0.06 |

Figure 3 shows the formation energy of isolated Cu ions (blue) and clusters (red) as a function of the number of atoms dissolved in the cell or clusters size. The linear dependence of the energy with the number of atoms in the case of isolated atoms indicates that interactions between Cu ions are negligible. The average energy required to move a Cu atom for the fcc metal into the silica matrix is 2.7±2.4 eV. We are unaware of prior experimental or theoretical values for this quantity but the predicted value can be compared with formation energy of a copper impurity in α-cristobalite $SiO_2$, which has been determined from DFT to be 4 eV [37]. As could be expected we find that the formation energy in the amorphous phase is lower than for the crystals where atoms have less freedom to adjust around the dissolved ion.

Importantly, the clusters are energetically favorable as compared to the isolated ions for all sizes studied. Figure 4 shows the clusterization energy, that is the energy difference between isolated copper atoms dissolved in silica and its cluster counter part, of Cu clusters with size 2-8 atoms. We also plot the clusterization energy for cluster with planar structure and single-atom chains of interest as the ultimate nanoscale conductive bridge. Unfortunately for ECM applications, the single atom chains are less stable than the corresponding equiaxed clusters indicating that such ideal filaments might be difficult to obtain in practice during the operation of ECM cells. This is in agreement with recent results from MD simulations [14]. Interestingly, we find that equiaxed clusters are energetically favorable over planar structures when the clusters are embedded in silica, the opposite trend as in vacuum. In addition, we show the formation energies of Cu clusters in vacuum; these results are in good agreement with prior DFT results [30, 31]. We note that the relative stability of clusters with respect to the isolated atoms is stronger in vacuum that in the silica matrix.

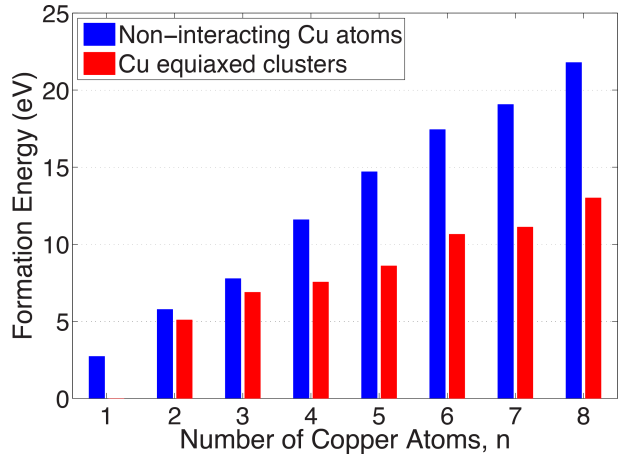

**Figure 3** Calculated formation energy of non-interacting Cu atoms and equiaxed Cu clusters embedded in a-SiO$_2$.

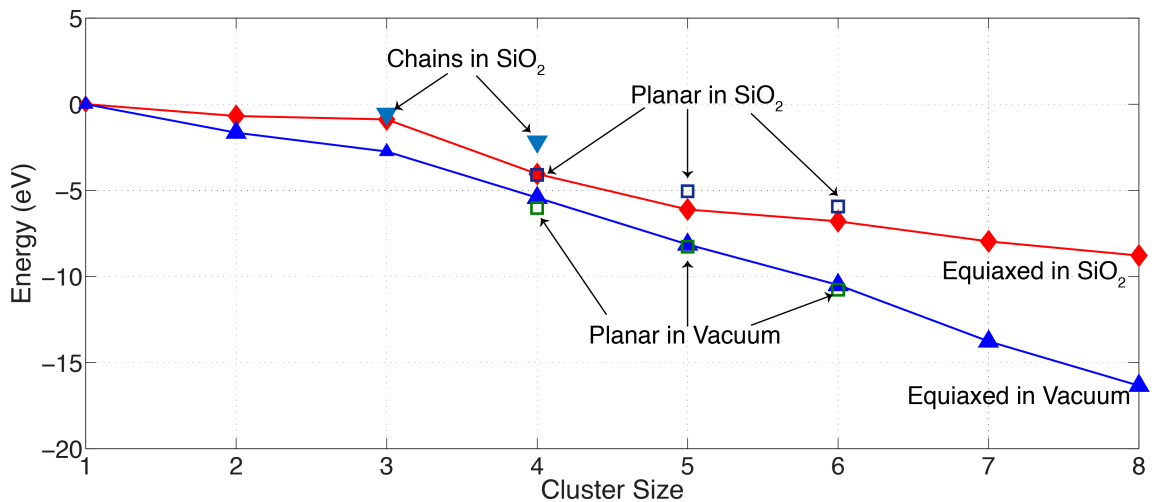

**Figure 4** Clusterization energy for small Cu clusters 1-8 atoms. Equiaxed copper clusters in silica are shown in red diamonds, equiaxed copper clusters in vacuum in blue triangle, Cu chains in dark blue triangles, planar structures in silica and vacuum are shown in purple and green open squares, respectively.

The partial pair correlation functions (or radial distribution functions) for Cu-Si (red) and Cu-O (blue) provide insight into the atomic arrangements around the Cu ions or clusters. Figure 5 shows these functions around Cu ions in equiaxed clusters ranging from 1 to 8 atoms in size. For all copper cluster sizes we observe that the first shell around Cu atoms consist of oxygen atoms at an average distance of ~1.8 Å. This is consistent with strong chemical reactivity between copper and oxygen. Interestingly, this first shell is rather broad indicating the disordered nature of the matrix. A sharp peak corresponding to Cu-Si is observed at approximately 3.2 Å. As expected, the pair correlation for Cu-O and Cu-Si decreases as the size of the copper cluster increases from 1 to 8 atoms.

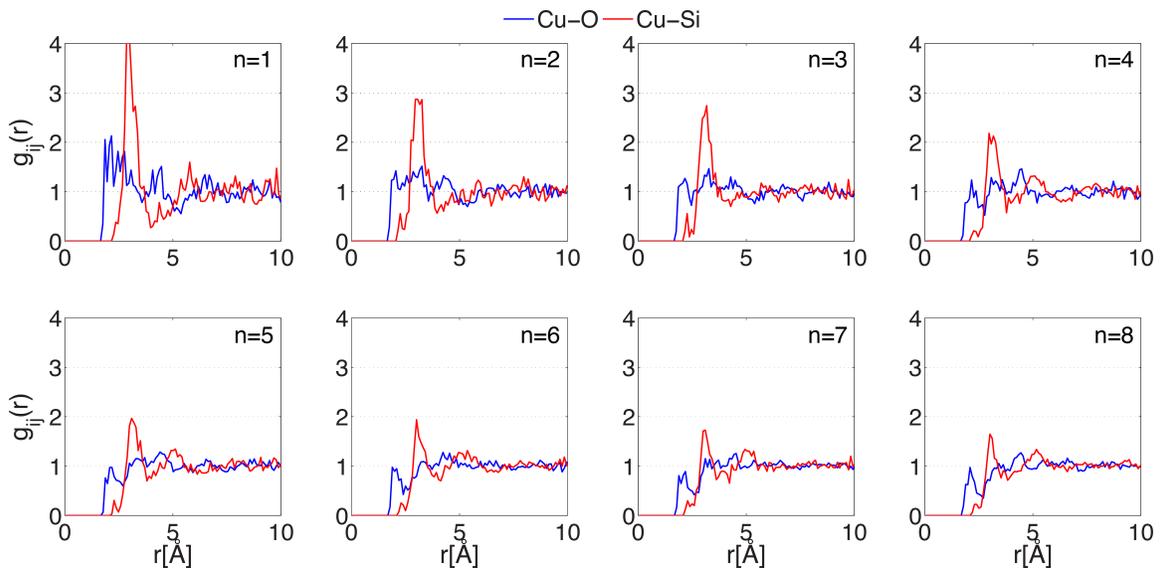

**Figure 5** Cu-Si (red) and Cu-O (blue) partial pair correlation functions of the $Cu_n/SiO_2$ system with n=1-6.

Understanding the ionicity of the Cu atoms and clusters is important to understand their bonding and their role in ECM cells. Calculations of the atomic charges via Bader's atoms in molecule method [38, 39] for the Cu clusters and chains are shown in Figure 6. The high electron transfer (the charge on the isolated Cu ions is close to +0.6 e) is consistent with strong Cu-O interaction also highlighted in the partial pair correlation functions. The calculations show that the average atomic charge is reduced by a factor of two for 5- and 6-atom clusters as compared with the single ion. This shows that relatively small clusters show some metallic character, which is important to understand the electrical conductivity of the nanoscale filaments in ECM cells. Interestingly, the average charge on the isolated Cu atoms decreased with increasing density of ions dissolved. Thus, even though the energetics shows little interactions between ions their electronic structure is changing. We note that the predicted Bader charges are significantly higher than the Mulliken populations reported previously, see supplementary material of Ref. [14]. We attribute this difference to the different methods used to extract it.

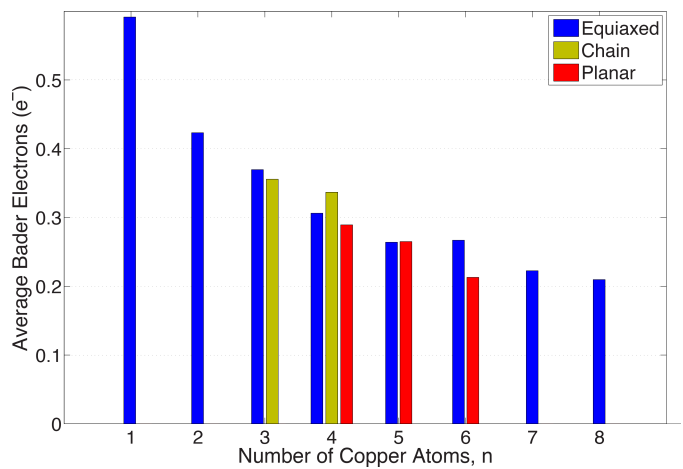

**Figure 6 Average Bader atomic charge of copper clusters (blue) and chains (red).**

## IV. ACTIVATION ENERGY FOR DIFFUSION

Figure 7(b) shows six representative energy landscapes for Cu diffusion in amorphous $SiO_2$. We note that in all but one cases the energy of the initial and final states (obtained from subsequent metastable states during a high-temperature MD simulation) represent local minima. This is a good validation of the approach to combine high-temperature MD simulations to identify metastable configurations with NEB at the DFT level.

We obtain a wide range of energy barriers associated with the migration of Cu ions through the amorphous network. Activation energies range from 0.2 eV to just over 1 eV with an average value of 0.60 eV obtained from NEB calculations over an ensemble of independent diffusion paths, see Figure 7(a). Experimental studies [40-42] show a strong dependence of diffusion on the material processing technique. Reported activation energy for diffusion of copper in amorphous silicon dioxide ranges from 0.7 eV observed in porous $SiO_2$ to 1.1 eV in thermal deposited silica. The spatial distribution of barriers is also important to understand ion migration and obtain an effective activation energy for diffusion; this is beyond the scope of this paper.

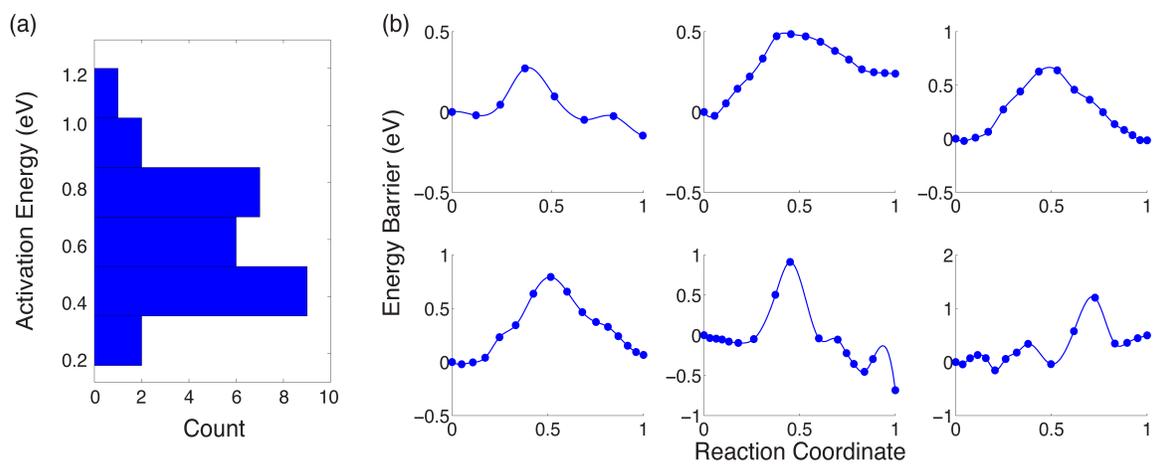

**Figure 7 (a)** Distribution of activation energy values of Cu in amorphous silicon dioxide obtained via transition state theory. **(b)** Representative minimum energy paths corresponding to the activation energies in (a)

## V. DISCUSSION

A fundamental knowledge of the dissolution, clusterization and migration of Cu in $SiO_2$ is important to understand and optimize ECM cells. However, the use of the results presented in this paper to predict performance of actual devices or interpret experiments should be done with care as it is well known that properties of ECM cells depend strongly on processing conditions and the resulting dielectric structure. For example results in Ref. [43] suggest a direct relationship between the threshold voltage and the annealing conditions for copper diffusion, leading to a modulation of the switching (threshold) voltage after each cycle. Thus, it is important to highlight that the results in our paper correspond to well annealed, non-porous amorphous silica. The atomistic structures have been shown to exhibit defects consistent with those observed experimentally [24, 25] and the relatively high cooling rates in the simulation result in higher defect densities compared to high-quality amorphous silica [24]

Experimental results on ECM memory devices [20, 44] suggest that the rate-limiting step for switching (SET/RESET) is the nucleation of stable copper cluster in $SiO_2$. Other reports [22, 44] proposed different filament growth and migration regimes depending on the ionic mobility of the amorphous dielectric matrix and the redox rates of the metallic inclusions. Importantly, our results show that the formation of copper cluster inclusions in amorphous silica is always energetically favorable and that no critical nucleus size exists. We stress that in our simulations the oxide is allowed to relax around the metallic clusters and, thus, the energetics presented here may not be applicable for ultra-fast switching devices.

## VI. CONCLUSIONS

We presented an atomistic study of the dissolution, clusterization and migration of Cu ions in amorphous silica matrix combining MD simulations to predict structures and DFT to refine them and predict the quantities of interest. Our results suggest that the clustering of copper atoms in silica is always energetically favorable if the matrix is able to relax around the metal cluster. Furthermore, we found that Cu atomic chains are substantially less stable compared to their cluster counterpart. The broad distribution of formation energies suggests a strong dependence on the local environment imposed by the amorphous matrix; this effect is also observed in the wide range of activation energies for diffusion obtained along different paths.

The results presented in this paper provide insight into the operation of ECM cells and quantitative values for fundamental materials properties that can be used in continuum models of these devices.

**ACKNOWLEDGEMENTS**

This work was supported by the FAME Center, one of six centers of STARnet, a Semiconductor Research Corporation program sponsored by MARCO and DARPA. Support by the US Department of Energy's National Nuclear Security Administration under Grant No. DE-FC52-08NA28617 is acknowledged. The computational resources from nanoHUB.org and Purdue University are also acknowledged.